\documentclass[conference]{IEEEtran}

\usepackage{epsfig, hyperref, amssymb, amsmath, amsthm, bbm, subcaption, xcolor,setspace}
\theoremstyle{definition} \newtheorem{definition}{Definition}

\DeclareMathOperator*{\argmax}{arg\,max}
\DeclareMathOperator*{\argmin}{arg\,min}

\begin{document}
\title{The Flow Fingerprinting Game}

\author{\IEEEauthorblockN{Juan A. Elices}
\IEEEauthorblockA{University of New Mexico\\
jelices@ece.unm.edu}
\and
\IEEEauthorblockN{Fernando P\'erez-Gonz\'alez}
\IEEEauthorblockA{University of Vigo\\
fperez@gts.uvigo.es}
}

\maketitle

\begin{abstract}
Linking two network flows that have the same source is essential in intrusion detection or in tracing anonymous connections. To improve the performance of this process, the flow can be modified (fingerprinted) to make it more distinguishable. However, an adversary located in the middle can modify the flow to impair the correlation by delaying the packets or introducing dummy traffic. 

We introduce a game-theoretic framework for this problem, that is used to derive the Nash Equilibrium. As obtaining the optimal adversary delays distribution is intractable, some approximations are done. We study the concrete example where these delays follow a truncated Gaussian distribution. We also compare the optimal strategies with other fingerprinting schemes.
The results are useful for understanding the limits of flow correlation based on packet timings under an active attacker.
\end{abstract}

\IEEEpeerreviewmaketitle

\section{Introduction}
Becoming anonymous is a goal for network attackers to avoid prosecution but also a necessity for dissidents, human rights activists, etc. This anonymity can be achieved by passing the traffic through a chain of relays. Network attackers generally use compromised hosts called stepping stones as relays~\cite{StHe:95}, while the rest of the users voluntary hosts that provide this service in low-latency anonymous networks. Interestingly, deanonymizing connections of these two kinds is essentially the same problem~\cite{ElPe:12}, which requires matching the egress and ingress flows. In these applications, the traffic is generally encrypted and sometimes divided into identical size packets, making the correlation of packet timings the most suitable solution.

Methods to find correlated flows can be classified in passive analysis and active watermarks. They differ in whether the flow is modified or not. Passive analysis needs slightly longer sequences but it can be less effective when timing patterns are very highly correlated, for instance two HTTP connections to the same web page. Watermarking schemes can avoid this problem but at the expense of being detectable~\cite{LuZhZhPeLe:11,LiHo:12}.

An adversary (AD), such as a stepping stone or an anonymous network relay, may modify the flow to prevent the correlation by introducing delays to packets or adding dummy packets to the flow. The existence of the AD has been considered in a passive analysis scenario by \cite{DoFlShPaCoStL:02} and \cite{BlSoVe:04}, where the AD is limited to delaying packets, and \cite{ElPe:13-2} analyzes a more complex AD model that, besides delaying packets, can also add and remove packets from the flow.

In this paper we study the limits of flow fingerprinting in an adversarial environment. Flow fingerprinting, which as flow watermarking, slightly perturbs the communication patterns, differs with the latter in that the modification is unique to each flow, so that every source sequence can be indistinctively identified.  To the best of our knowledge, the only active flow fingerprint method is Fancy~\cite{HoBo:13}. 

To overcome the loop of proposing an attack and creating an \emph{ad-hoc} solution, we propose a game-theoretic framework and look for the optimum strategies that the players, traffic analyst (TA) and AD, should adopt. A similar game-theoretic framework has been used in other contexts such as Information Hiding~\cite{MoOs:03}, Source Identification~\cite{BaTo:13} or in passive traffic analysis~\cite{ElPe:13-2}.

The rest of the paper is organized as follows: in Section \ref{sec:not} we introduce the notation, together with some basic concepts of game theory. Section \ref{sec:mod} presents a rigorous definition of the flow fingerprinting game. In Section \ref{sec:det} we derive the used detector. Section \ref{sec:gau} studies the case when the attack channel is distributed as a truncated Gaussian. Section \ref{sec:sim} validates the performance and each player decisions using a simulator, and also presents a comparison of our scheme with Fancy in terms of error probability. Conclusions are presented in Section \ref{sec:con}.

\section{Notation}\label{sec:not}
We use the following notation. Random variables are denoted by capital letters (e.g., $X$), and their individual realizations by lower case letters (e.g., $x$). 
The domains over which random variables are defined are denoted by script letters (e.g., $\mathcal{X}$). Sequences of $n$ random variables are denoted with $X^n$ if they have random nature or by $x^n$ if they are deterministic. $X_i$ or $x_i$ indicate the $i$−th element of $X^n$ or $x^n$, respectively. The  probability distribution function (pdf) of a random variable $X$ is denoted by $f_X(x),\;x \in\mathcal{X}$. We use the same notation to refer to pdf of sequences, i.e. $f_{X^n}(x^n),\;x^n \in\mathcal{X}^n$. When no confusion is possible, we drop the subscript in order to simplify the notation. We denote with $\Delta$ the difference operation of a sequence, i.e $\Delta x^n=\{x_2-x_1, \dots, x_n-x_{n-1}\}$ and with $\Delta x_i=x_{i+1}-x_i$ the $i$th element of this sequence.

\subsection{Performance Metrics}

To measure performance, we use two metrics: the probability of detection ($P_D$) and the probability of false positive ($P_F$). Given two hypotheses: $H_0$ and $H_1$, $P_D$ is the
probability of deciding $H_1$ when $H_1$ holds, whereas $P_F$ is the probability of deciding $H_1$ when $H_0$ holds.

Typically, performance is graphically represented using the so-called
ROC (Receiver Operating Characteristic) curves, which represent
$P_D$ vs. $P_F$. In order to compare different ROCs in a simple way, we use the AUC (area under the ROC curve), a measure that takes a value of 1 in the case of perfect detection and 0.5 in the case of random choice. 

\subsection{Game Theory}

Game theory is the mathematical study of interaction among intelligent rational decision-makers. Formally, a two player game is defined as a quadruple $G(A_1,A_2, u_1, u_2)$, where  $A_i= \{a_{i,1},\dots a_{i,n_i}\}$ are the actions available to the $i$ player, $u_i: A_1 \times A_2 \mapsto \mathbb{R}, \; i = 1,2$ is the utility function or payoff of the game for player $i$. An action profile is the double $a \in A_1 \times A_2$.
We  are interested in zero-sum games, where $u_1(a)+ u_2(a)=0, \forall a \in A_1 \times A_2$, which means that the gain (or loss) of utility of player 1 is exactly balanced by the losses (or gains) of the utility of player 2. In this case, we can simplify the game notation to a triplet $G(A_1,A_2, u)$, where $u=u_1=-u_2$. 

We say that an action profile $(a_{1,i^*}; a_{2,j^*})$ represents a Nash
equilibrium (NE) if 
\begin{align}
u(a_{1,i^*}; a_{2,j^*}) \geq u(a_{1,i}; a_{2,j^*})\;\; \forall a_{1,i} \in A_1\nonumber\\
u(a_{1,i^*}; a_{2,j^*}) \leq u(a_{1,i^*}; a_{2,j})\;\; \forall a_{2,j} \in A_2,
\end{align}
intuitively this means that none of the players can improve his utility by modifying his 
strategy assuming the other player does not change his own.

Games can be classified in simultaneous games, where both players move unaware of the other player action, and sequential games, where later players have some knowledge about earlier actions. In sequential games, an action profile is a subgame perfect equilibrium (SPE) if it represents a NE of every subgame of the original game. Therefore, a SPE  is a refinement of the NE that eliminates non-credible threats.

\section{Flow Fingerprinting Game}\label{sec:mod}

\begin{figure}
  \centering
    \includegraphics[width=0.85 \columnwidth]{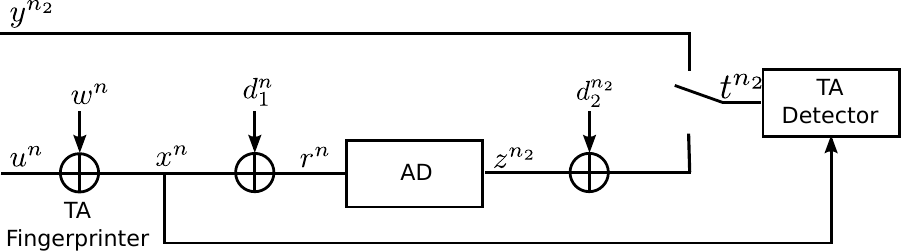}
  \caption{Model of the FFG}
  \label{fig:ffg}
\end{figure}

The Flow Fingerprinting Game (FFG) is represented in Figure \ref{fig:ffg}. In this game, there are two players: the Traffic Analyst (TA) and the Adversary (AD).
 
The task of the TA is to accept or reject the hypothesis that a flow $t^{n_2}$ is indeed the same flow as a known one, $u^n$. To improve the efficiency, the TA can modify the flow  by embedding a fingerprint $w^n$. Due to the nature of the problem the modification must be additive, i.e., $x^n=u^n+w^n$. We constraint the fingerprint to delay any packet at most $W_C$ seconds. This flow suffers a network delay of $d_1^n$ before reaching the AD. We denote by $r^n$ the flow received by the AD.

The goal of the AD is to modify the flow, producing $z^{n_2}$, in such a way that the detector decide that this sequence is not related with $u^n$.  In order to do this, the AD can delay any packet at most $A_C$ seconds and add up to $P_A\cdot n$ dummy packets, hence $P_A$ is the maximum ratio between chaff and real traffic. We denote by $a^n$ the sequence of delays inserted by the AD to each packet, and by $c^{n_A}$ the chaff packets timing sequence, which consists of $n_A$ packets. The output flow of the AD $z^{n_2}$ suffers an additional delay $d_2^{n_2}$ due to the network between the AD and the detector.

We represent by $D$ the delay suffered by a packet in the whole path, i.e. $D=D_1+D_2$. Note that $\Delta D$ is the packet delay variation (PDV), also called jitter.

Let $Y^{n_2}$ represent flows without any relation to $x^n$, but from the same application, and we assume that $f_{\Delta Y}(\Delta y)$ is known by both players and define the hypotheses: 
\begin{align*}
H_0: &\; t^{n_2} \text{ is not a fingerprinted version of }x^n\\
H_1: &\; t^{n_2} \text{ is a fingerprinted version of }x^n.
\end{align*}

We define the FFG as follows:
\begin{definition} 
The $FFG(A_{TA};A_{AD}; u)$ is a simultaneous, zero-sum game played by the TA and the AD, where
\begin{itemize}
\item The set of actions the TA can choose from, i.e. $A_{TA}$, is the duple of possible fingerprint sequences $w^n$ and acceptance regions $\Lambda_1$ for which $P_F$ is below a certain threshold $\eta$:
\begin{align}\label{eq:rfp}
A_{TA}=&\{w^n \times \Lambda_1: 0\leq w_i \leq W_C, \forall i\in [1,n],\nonumber\\ &Pr(Y^m \in \Lambda_1)<\eta\}
\end{align}

\item The set of possible attacks that the AD can choose from
\begin{align}
A_{AD}=&\{f(z^{n_2}|r^n): \exists (a^n, c^{n_A})\, |\, \forall_i \in [1,n],\, 0\leq a_i \leq A_C; \nonumber \\ & \frac{n_A}{n} \leq P_A;\, \forall_j \in [1,n_A], 0 \leq c_j \leq r_n+A_C; \nonumber \\ & 
 z^{n_2}= \text{sort}((r^n+a^n) || c^{n_A}) \},
\label{eq:sad1}
\end{align}
where $||$ represents the concatenation of sequences, and sort($x^n$) is a function that returns a sorted version of the input sequence.
\item The utility function is $P_D$, namely:
\begin{equation}
u(A_{TA}, A_{AD})= Pr( T^{n_2} \in \Lambda_1|H_1)
\end{equation}
\end{itemize}
\end{definition}

\subsection{Subgame Perfect Equilibrium}
As the players choose their actions in a given order, then the SPE needs to assume that a given player knows which actions have taken place before his own (otherwise the player could improve its utility given this information). Hence the solution to the game is: 
\begin{equation}\label{eq:mim}
u=\max\limits_{w^n} \min\limits_{A_{AD}}\max\limits_{\Lambda_1}u(A_{TA},A_{AD}).
\end{equation}.

As correlating directly timing sequences needs a precise estimation of $f_D$ and in a real implementation this is difficult to obtain, the use of the difference timing sequence, known as inter-packet delays (IPDs), seems more reasonable and has been widely adopted in the literature~\cite{HoBo:13, DoFlShPaCoStL:02, ElPe:13-2}.
The optimal detector, according to Neyman-Pearson Lemma, is the likelihood ratio test:
\begin{align} \label{eq:det1}
&\Lambda_1(t^{n_2},x^n, \hat{f}(z^{n_2}|r^n))= \int_{\mathcal{R}^n}\int_{\mathcal{Z}^{n_2}} \frac{ f_{\Delta D_2^{n_2}}(\Delta (t^{n_2}-z^{n_2}))}{f_{\Delta Y^{n_2}}(\Delta t^{n_2})}\nonumber \\
&\cdot \hat{f}(z^{n_2}|r^n) f_{\Delta D_1^n} (\Delta (r^n-x^n)) dz^{n_2} d r^n.
\end{align}
where $\hat{f}(z^{n_2}|r^n)$ is the assumed distribution of $f(z^{n_2}|r^n)$ by the detector.
The test chooses $H_1$ whenever $\Lambda_1(t^{n_2},x^n, \hat{f}_{Z^{n_2}|R^n})\geq \epsilon$, where $\epsilon$ is a threshold chosen to achieve $P_F<\eta$. At the SPE, $\hat{f}(z^{n_2}|r^n)=f(z^{n_2}|r^n)$ that gives a utility of
\begin{equation}\label{eq:mim2}
u=\max\limits_{w^n} \min\limits_{f(z^{n_2}|r^n)} Pr(\Lambda_1(T^{n_2},u^n+w^n, f(z^{n_2}|r^n))>\epsilon).
\end{equation}
Unfortunately, solving \eqref{eq:mim2} is a computationally intractable problem. In the following sections, we try to approximate the SPE by simplifying the problem.

\section{Detector}\label{sec:det}
In this section we derive a detector that is implemented in two steps: first, a matching process takes place that outputs two sequences of the same size, and then a likelihood test that needs a one-to-one correspondence between the flows is constructed. 

\subsection{Matching Process}
When dummy packets are added, i.e. $P_A>0$, there does not exist a one-to-one relation between the flows $x^n$ and $t^{n_2}$. To deal with this problem, we match each packet of $x^n$ with the most likely from $t^{n_2}$, later removing those packets of $t^{n_2}$ that have no correspondence on $x^n$. We denote this matching process as $t^n=m(x^n, t^{n_2})$.

We represent the fact that the $i$th packet from $x^n$ is paired with the $j$th packet from $t^{n_2}$ by $p(i)=j$. Let ${\mathcal M}$ be the set of all injective functions from $\mathcal{N}=\{1,\dots,n\}$ to $\mathcal{N}_2=\{1,\dots,n_2\}$, i.e $\forall i_1, i_2 \in \mathcal{N}, \; p(i_1)=p(i_2) \implies i_1=i_2$. Then the matching function $m(x^n, t^{n_2})$ is the function from $\mathcal M$ that minimizes the mean square error between $x^n$ and a shifted version of $t^{n_2}$ as follows:
\begin{equation}
m=\argmin_{\mathcal{M}}\sum_{i=1}^n (t_{p(i)}-x_i-\rho-E(a_i))^2
\end{equation}
where $E(a_i)$ is the expected value for the delay added by AD to the $i$th packet, recall that $f(a^n|x^n)$ is assumed to be known by the detector, and $\rho$  is a synchronization constant equal to the sample mean of the delays, i.e. $\rho=\frac{1}{n} \sum_{i=1}^n d_i$. In a real implementation, where the sample mean is unknown, $\rho$ can be obtained through an exhaustive search (self-synchronization property).

\subsection{Likelihood Test}
Confining the detector to those based on first-order statistics of the IPDs for feasibility reasons, the optimal likelihood ratio test becomes:
\begin{align} \label{eq:det1}
\Lambda_1(t^n,x^n,f(a^n|\Delta x^n))=&\sum_{i=1}^{n-1}  \bigg(\iint_{\mathcal{A}^2}\frac{f_{\Delta D}(\Delta t_i -\Delta \hat{a}_i)}{f_{\Delta Y}({\Delta w_i})}\nonumber\\ &\cdot f_{A_{i,i+1}|\Delta X^n}(\hat{a}_{i,i+1}|\Delta x^n)d\hat{a}_i d\hat{a}_{i+1}
 \bigg),
\end{align}
where
\begin{equation}
f(a^n|\Delta x^n)=\int_{\mathcal{R}^n}f(a^n|\Delta r^n)f_{\Delta D_1}(\Delta (r^n-x^n))dr^n.
\end{equation}

\section{Truncated-Gaussian Attack Channel}\label{sec:gau}

As finding the distribution $f(a^n| \Delta r^n)$ that minimizes $Pr(\Lambda_1 (t^{n},x^n, f(a^{n}| \Delta  r^n))>\epsilon)$ may not be feasible for the AD, we study the case when the distribution of the delays introduced by the AD is constrained to a truncated Gaussian in the interval $[0,A_C]$. Hence, $a_i \sim N(\mu_i, \sigma^2 | 0 \leq a_i \leq A_C)$. Note that this model includes as limits the deterministic attack $\sigma^2 \to 0$, and a uniform attack $\sigma^2 \gg A_C^2$. 

The AD selects the sequence of means $\mu^n$ and the variance $\sigma^2$, as well as the timing for the dummy packets, hence $A_{AD}=\{\mu^n \times \sigma^2 \times c^{n_A}\}$.

From \eqref{eq:mim}, the SPE actions, denoted with the superscript *, are
\begin{align}
a^*_{TA}&=\argmax_{w^n} \min_{A_{AD}} Pr(\Lambda_1(m(T^{n_2},x^n),x^n, f(a^n|x^n)) > \epsilon)\\
a^*_{AD}&=\argmin_{A_{AD}} Pr(\Lambda_1(m(T^{n_2},x^n),x^n, f(a^n|x^n)) > \epsilon)\label{eq:ad1},
\end{align}
where
\begin{equation}
Pr(\int_{\mathcal{X}^n}\Lambda_1(m(Y^{n+\lfloor P_A \cdot n\rfloor },x^n),x^n,f(a^n|x^n))dx^n \leq \epsilon)  = \eta. \label{eq:pfp}
\end{equation}
As the AD must decide its action in real time and \eqref{eq:ad1} is computationally expensive, we approximate each decision individually, as explained next
\subsubsection{Mean sequence ($\mu^n$)} a good approximation when $\Delta D_2$ has zero mean and its variance is much smaller than $\Delta Y$ (as it is the case in practice) is
\begin{equation}
(\mu^{n})^* \approx \argmin_{\mathcal{\mu}^n} \sum_{i=1}^{n-1} \log f_{\Delta Y}(\Delta r_i+ \Delta \mu_i)\label{eq:apmu}.
\end{equation} 
Note that under this approximation the AD is maximizing the likelihood of $r^n+a^n$ coming from $y^n$, i.e., making the sequence as typical as possible.

\subsubsection{Variance ($\sigma^2$)}
The value of $\sigma^2$ presents a trade-off: small values of $\sigma^2$ make $a^n$ to be chosen so that the sequence looks more similar to the typical sequence of $Y^n$ but with the disadvantage that the uncertainty of $a^n$ for the detector is smaller. Recall that the detector is assumed to know $f(a^n|\Delta x^n)$.

\begin{figure}
  \centering
    \includegraphics[width=0.78 \columnwidth]{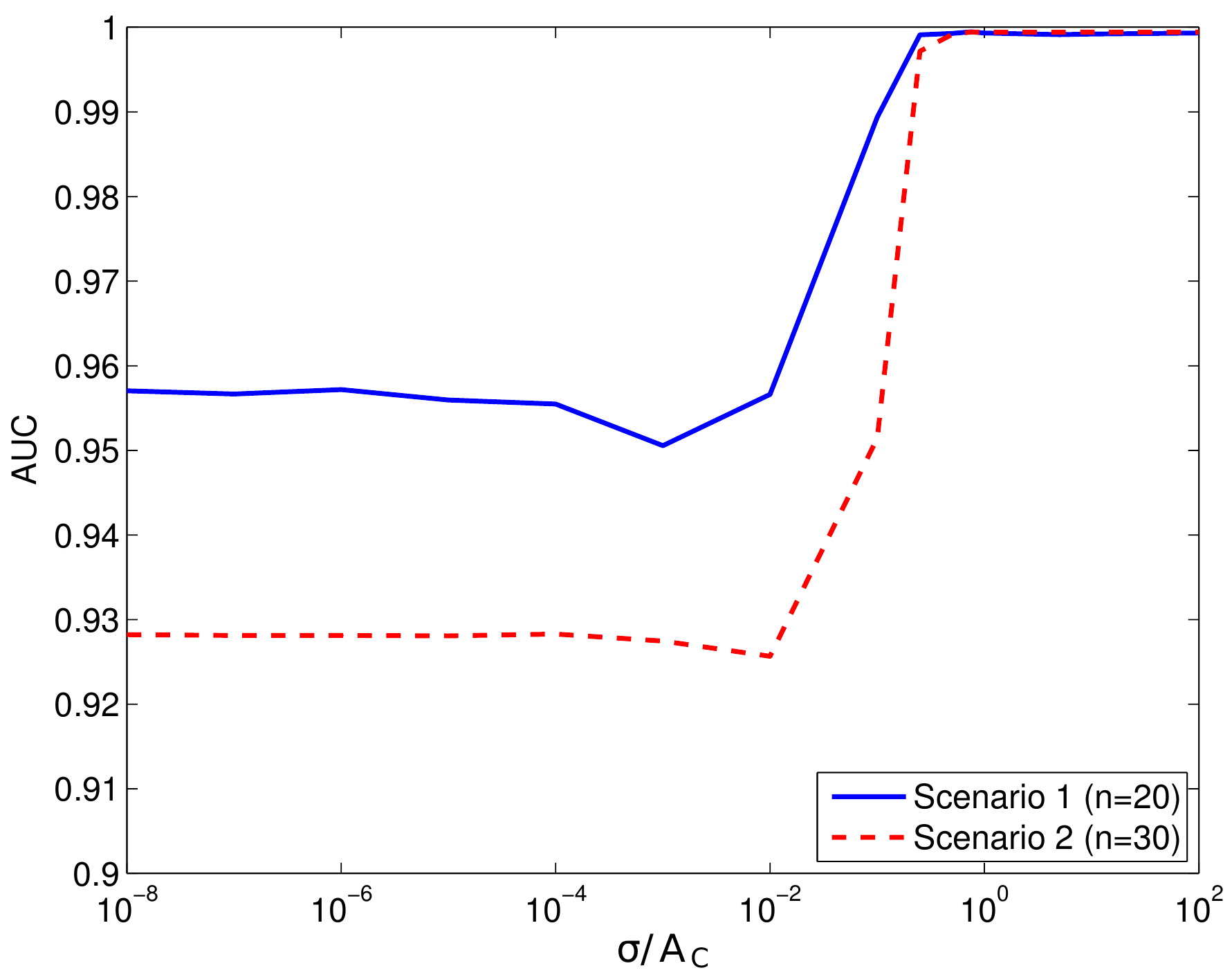}
  \caption{Dependence of the performance with $\sigma$ ($A_C=250$ms,  $P_A=0$, 
$W_C=0$ms).}
  \label{fig:VarAUC}
\end{figure}

\begin{figure}
  \centering
	\includegraphics[width=0.78\columnwidth]{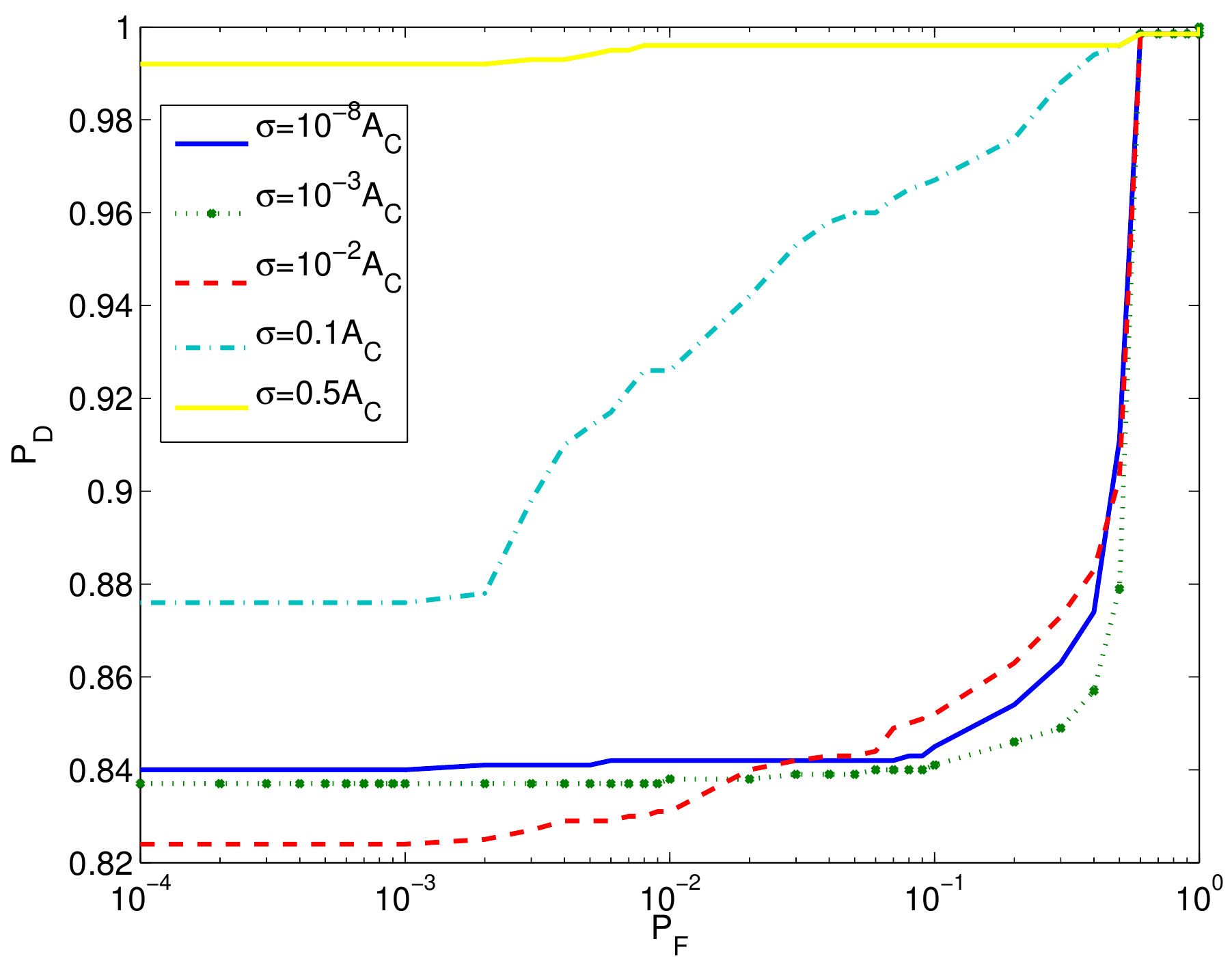}
  \caption{ROCs for different variances in Scenario A ($n=20$, $A_C=250$ms,  $P_A=0$, 
$W_C=0$ms).}
	\label{fig:VarRocSc1}
\end{figure}

We calculate the value of $\sigma^2$ that minimizes \eqref{eq:ad1} empirically using the simulator and the scenarios presented in next section. A graph of the variation of AUC with $\sigma$ is depicted in Figure \ref{fig:VarAUC}. The constant region on each side corresponds to a deterministic ($\sigma \to 0$) and to a uniform attack ($\sigma \gg A_C$). The minimum performance is obtained in the interval $[10^{-3}, 10^{-2}] \cdot A_C$. This small variance makes the attack virtually deterministic ($\sigma \to 0$), implying that making the sequence more typical is the prevailing factor. Figure \ref{fig:VarRocSc1} let us see better the difference of performance for different values of $\sigma$. In the following unless otherwise specified, the AD chooses $\sigma = 10^{-2}\cdot A_C$.

\subsubsection{Chaff traffic ($c^{n_A}$)}
Assuming that $f_{\Delta D} (\Delta d)$ is symmetric around its mean and unimodal (as it is the case in practice), then the matching process selects those packets that give a higher value of $\Lambda_1(t^n,x^n,f(a^n|\Delta x^n))$. Therefore, the AD will choose $c^{n_A}$ so that these dummy packets are removed in the matching process. On the other hand, the AD will need these packets to force the TA to consider longer possible sequences for $y^{n+n_A}$. According to \eqref{eq:pfp}, longer sequences of $Y^{n_2}$ will increase $\epsilon$.

\subsubsection{Fingerprint ($w^n$)}

This is the converse problem as the delays for the AD. The TA wants $x^n$ to be as distinguishable as possible from the typical sequence of $y^n$. Then,
\begin{equation}
(w^{n})^* \approx \argmax_{w^n} \sum_{i=1}^{n-1} \log f_{\Delta Y}(\Delta u_i+ \Delta w_i)\label{eq:fing}.
\end{equation}

\section{Performance}\label{sec:sim}
In this section we present the two scenarios we use in the remaining of the paper and construct a simulator. Afterwards, we compare the performance modifying one action at a time: the detector; the AD action, and the fingerprinting scheme. This will show the impact of those actions on the utility.

\subsection{Scenarios and Simulator}

We present two scenarios, A and B, that we use in the sequel to evaluate the performance. Scenario A represents a stepping stone that forwards SSH traffic inside the Amazon Web Services network. The TA-Fingerprinter, the AD and the TA-Detector (cf. Figure \ref{fig:ffg}) are EC2 instances located in Virginia, Oregon and California, respectively.  We use the IPDs from $8746$ replayed SSH connection captures with 64 million packets from \cite{KoHeAbYe:04} and \cite{BaSaPrVa:10}. The simulated delays correspond to Scenario 10 from \cite{ElPe:13}. 

Scenario B simulates a web page accessed from the Tor network whose real origin is to be found. In it, the TA-Fingerprinter corresponds to the web server, the AD to the Tor entry relay, and the TA-Detector to the client.  We use the IPDs of $113690$ replayed HTTP connections that sum around 139 million packets taken from the same repositories. The delays correspond to the measurements of Scenario 11 from \cite{ElPe:13}.

The calculation of $\Lambda_1$ in \eqref{eq:det1} needs an estimation of $f_{\Delta D}$ and $f_{\Delta Y}$. To this end we apply kernel smoothing techniques~\cite{BoAz:97}. As it is customary, we separate the data gathered for each scenario into two subsets: training, to estimate the pdfs, and test, used in the simulator, assigning 50\% of the samples to each.

Simulations are carried out in the following way. First, we generate a timing information from the measured IPDs (as explained in the following paragraphs), $u^{n}$. Then we introduce the fingerprint $w^n$ according to \eqref{eq:fing} obtaining $x^n$. Afterwards, $d_1^n$ is added to each packet of $x^n$ using the measured delays, obtaining $r^{n}$. Next, we generate $a^n$ and $c^{n_A}$ according to the procedure discussed in Section \ref{sec:gau}, obtaining $z^{n_2}$. Subsequently, we introduce another delay $d_2^n$ to generate sequence $t_1^{n_2}$.

We generate a second timing sequence $y^{n_2}$ with $n_2=n+\lfloor P_A \cdot n\rfloor$. This sequence has the purpose of evaluating the performance under $H_0$. Finally, we use the test from \eqref{eq:det1} to obtain both $\Lambda_1(t^{n_2}_1, x^n, f(a^n|x^n))$ and $\Lambda_1(y^{n_2}, x^n, f(a^n|x^n))$. This experiment is repeated $10^4$ times when the simulation has an AD or $10^5$ when no AD is present. For different values of $\epsilon$ we obtain $P_D$ as the rate of $\Lambda_1(t^{n_2}_1, x^n, f(a^n|x^n))>\epsilon$, and $P_F$  as the rate of $\Lambda_1(y^{n_2}, x^n, f(a^n|x^n))>\epsilon$. 

Sequences $u^n$ and $y^n_2$ are generated in the following way: we place all the IPDs from the test set on an order-preserving list. The starting point is randomly selected from the list and the generated IPDs are the following values. 

The provided delays are sampled each $50$ms. We select one value randomly that we consider time 0 ms; the following values represent the delay at times 50 ms, 100 ms and so on. To obtain the delays at times where we do not have a measurement, we use linear interpolation.

\subsection{Detector comparison}
\begin{figure}
  \centering
  \begin{subfigure}[b]{\columnwidth}
	\centering
	\includegraphics[width=0.77\textwidth]{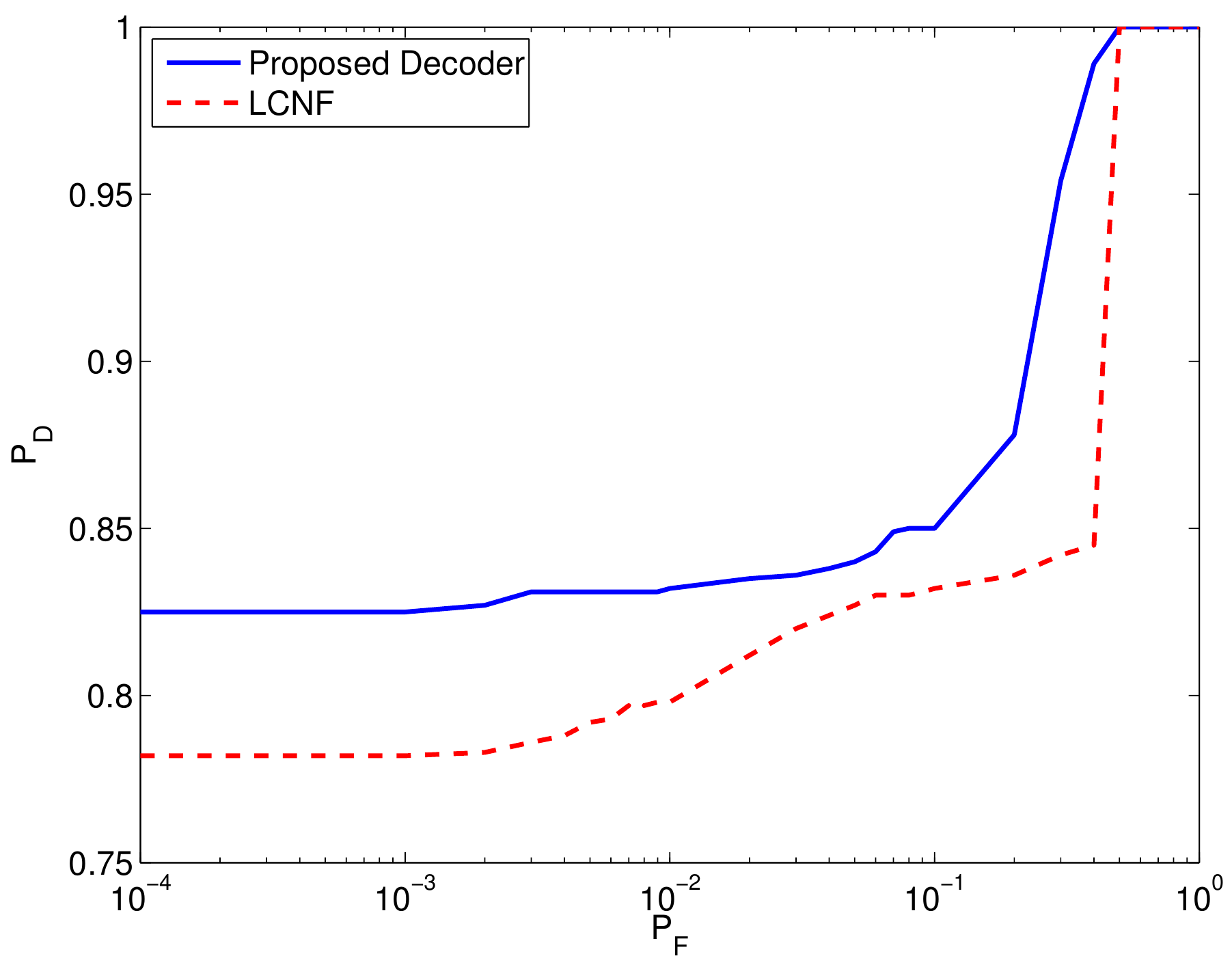}
	\caption{Scenario A ($n=20$, $A_C=250$ms, $P_A=1$, $W_C=0$ms).}
	\label{fig:CompDecSc1}
  \end{subfigure}
  \begin{subfigure}[b]{\columnwidth}
	\centering
	\includegraphics[width=0.77\textwidth]{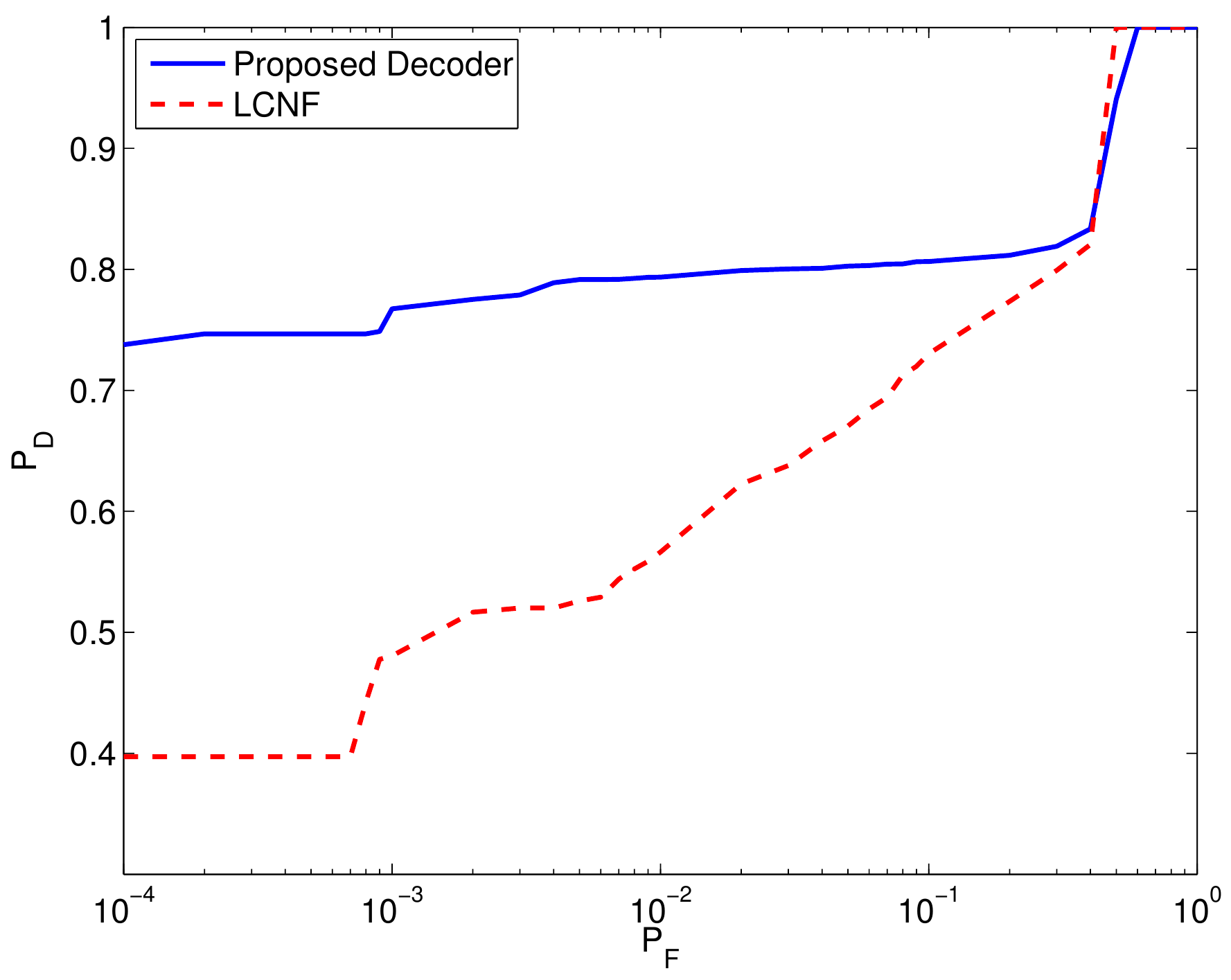}
	\caption{Scenario B ($n=30$, $A_C=250$ms, $P_A=1$, $W_C=0$ms).}
	\label{fig:CompDecSc2} 
  \end{subfigure}
  \caption{Comparison of the proposed detector with LNCF.}
\end{figure}

We compare our detector with the one used in \cite{ElPe:13-2} that we denote as LCNF (Linking Correlated Network Flows). This detector is claimed to be the optimal among those that estimate a value for $a^n$ and compensate it. Results are depicted in Figures \ref{fig:CompDecSc1} and \ref{fig:CompDecSc2} for Scenarios A and B, respectively. We see that our detector outperforms LCNF in both scenarios. Note that our detector is derived to be optimal among those which use just first-order statistics. Hence, by using higher-order statistics the performance could be improved at the expense of a higher computational cost.

\subsection{AD actions}
\begin{figure}
  \centering
  \begin{subfigure}[b]{\columnwidth}
	\centering
	\includegraphics[width=0.77\textwidth]{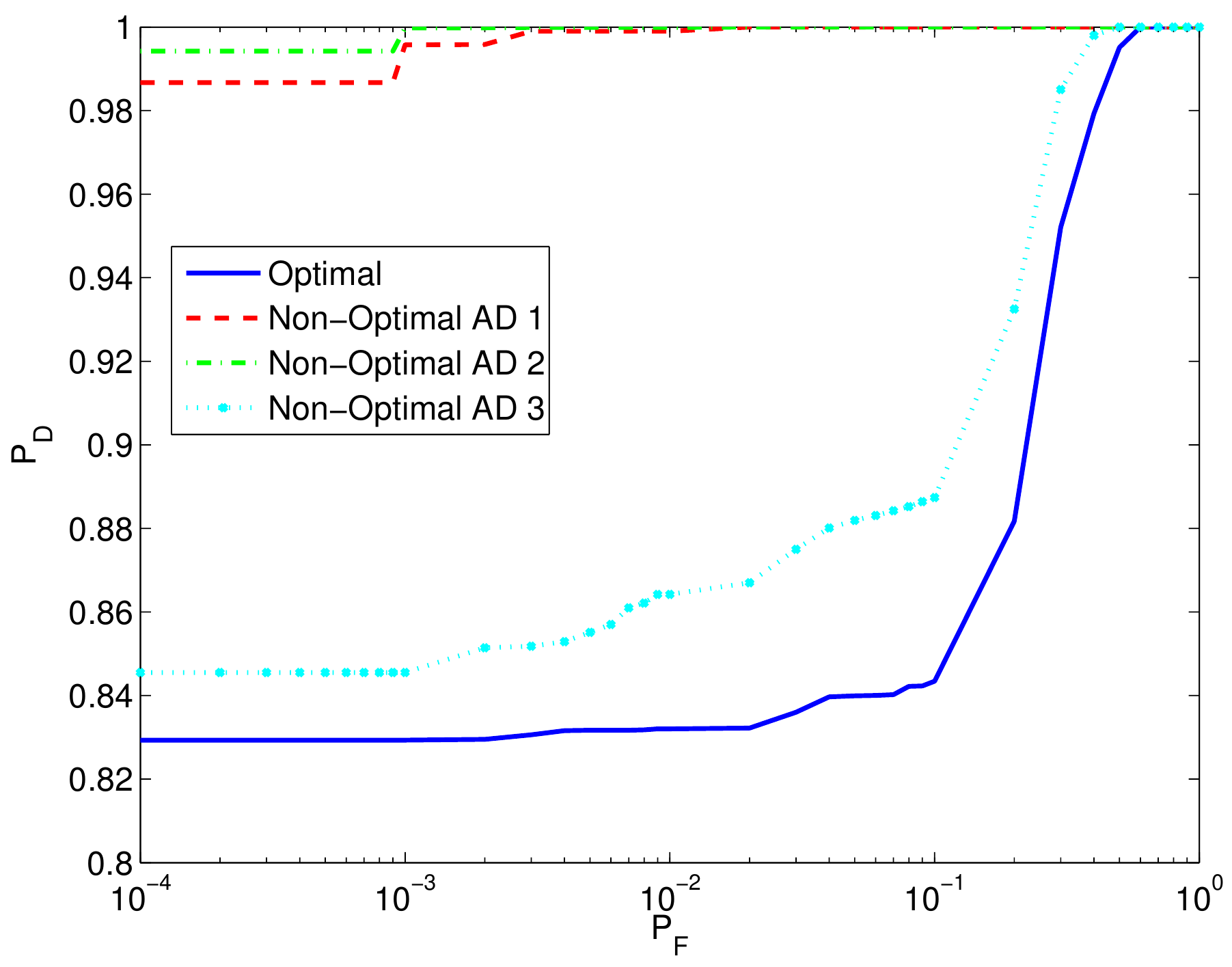}
	\caption{Scenario A ($n=20$, $A_C=250$ms, $P_A=1$, $W_C=0$ms).}
	\label{fig:Sc1CompAt}
  \end{subfigure}
  \begin{subfigure}[b]{\columnwidth}
	\centering
	\includegraphics[width=0.77\textwidth]{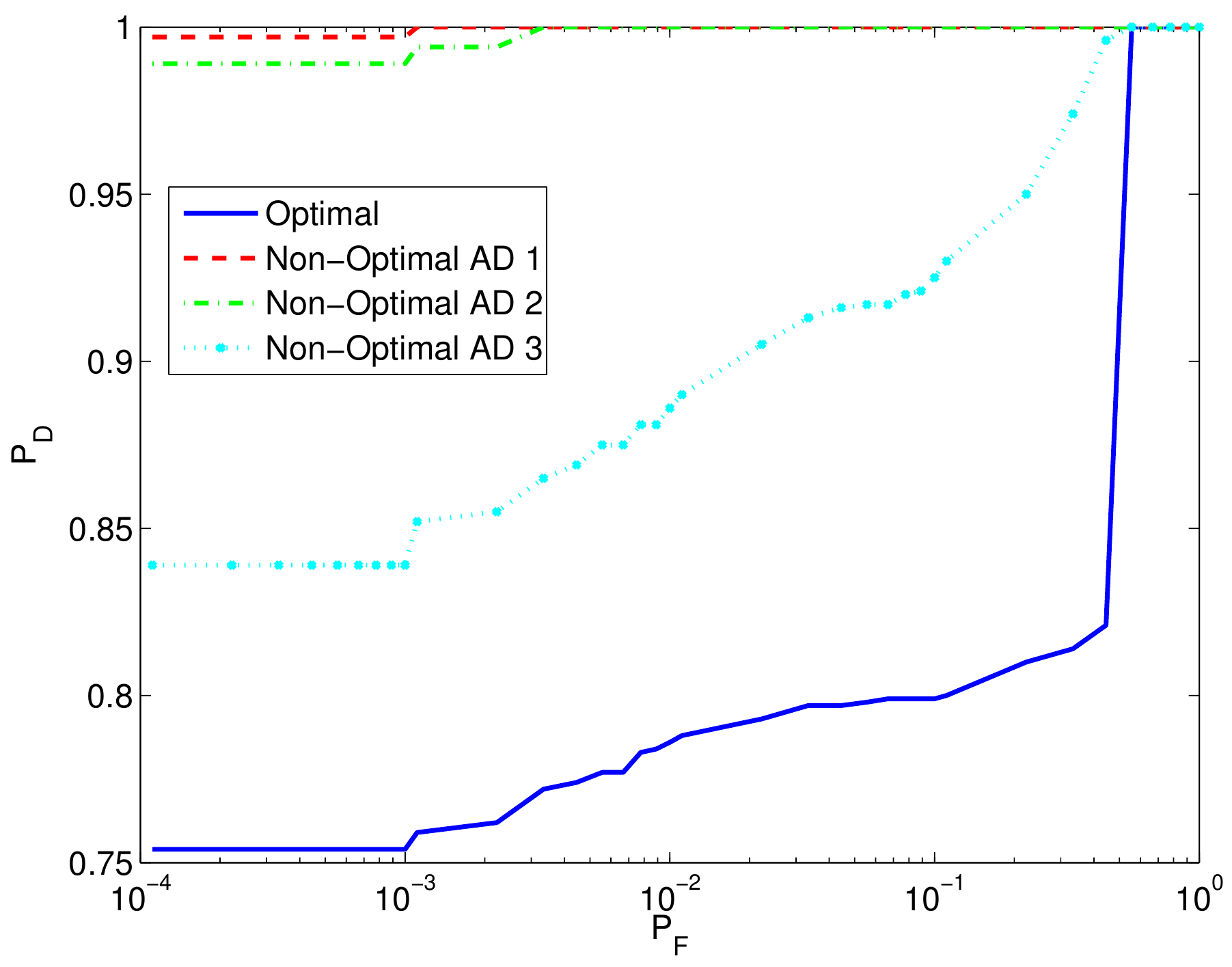}
	\caption{Scenario B ($n=30$, $A_C=250$ms, $P_A=1$, $W_C=0$ms).}
	\label{fig:Sc2CompAt} 
  \end{subfigure}
  \caption{Comparison of different AD strategies}
\end{figure}

We compare the ROC under an optimal adversary with those corresponding to three non-optimal adversaries: a) the AD selects $\mu_i$ randomly according to a uniform distribution between 0 and $A_C$; b) the AD chooses its delays from a uniform distribution; c) the AD chooses its delays $a^n$ as explained in Section \ref{sec:gau} but the chaff traffic is selected randomly, i.e. $c^{n_A}$ is an i.i.d. sequence uniformly distributed between the timing of the first packet and the last one. Results are depicted in Figures \ref{fig:Sc1CompAt} and \ref{fig:Sc2CompAt}. The conditions used are $A_C=250$ms, $P_A=1$, $W_C=0$ms, $n=20$ in Scenario A, and $n=30$ in Scenario B. We see that the delay distribution has a great effect on both scenarios but the dummy packets have a more significant influence on Scenario B. 
In any case, notice that the AD attack derived in Section \ref{sec:gau} impairs the flow correlation in a much more severe way than the suboptimal strategies.

\subsection{Fingerprint actions}
\begin{figure}
  \centering
  \begin{subfigure}[b]{\columnwidth}
	\centering
	\includegraphics[width=0.77\textwidth]{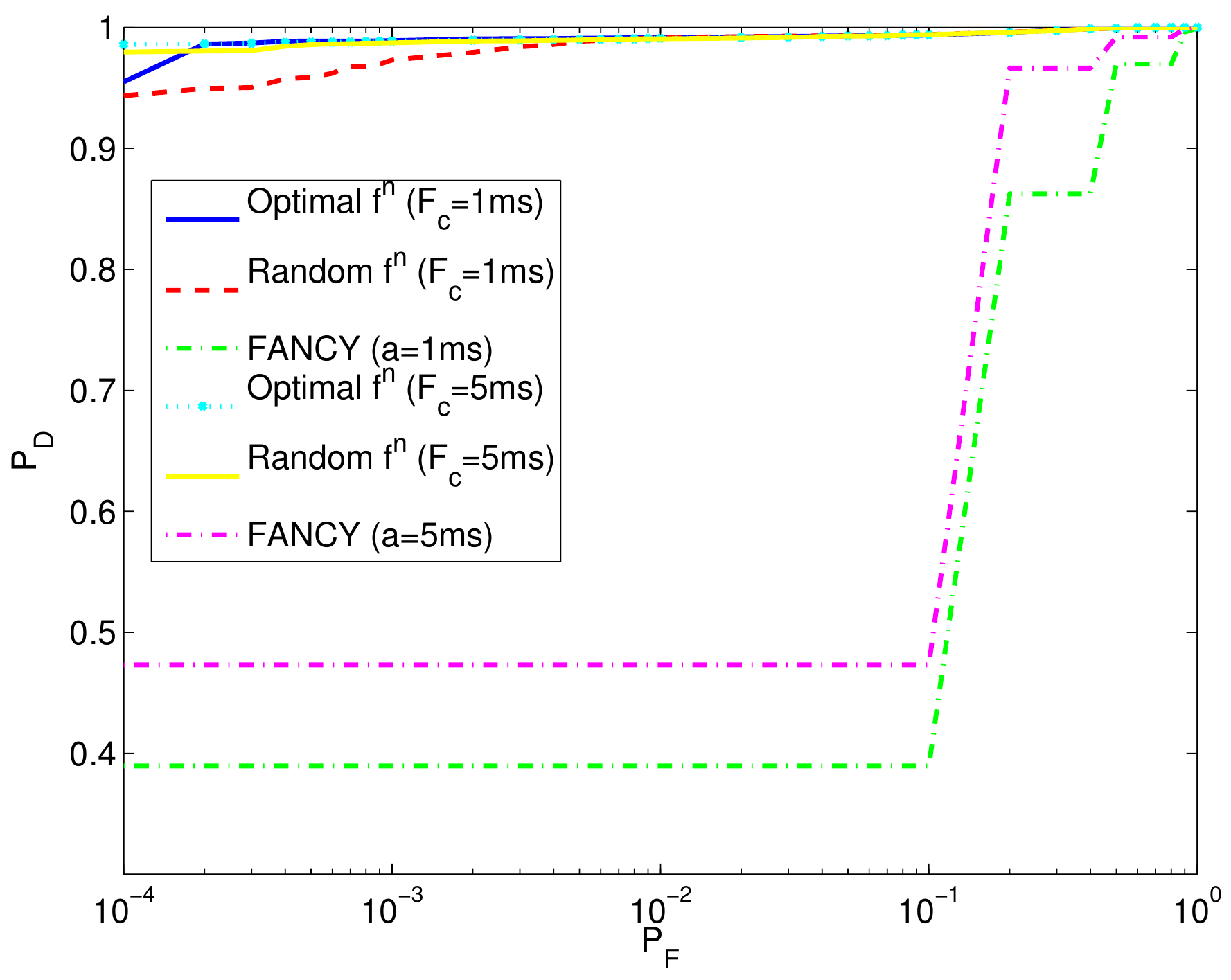}
	\caption{Scenario A ($n=5$).}
	\label{fig:Sc1FNAD}
  \end{subfigure}
  \begin{subfigure}[b]{\columnwidth}
	\centering
	\includegraphics[width=0.77\textwidth]{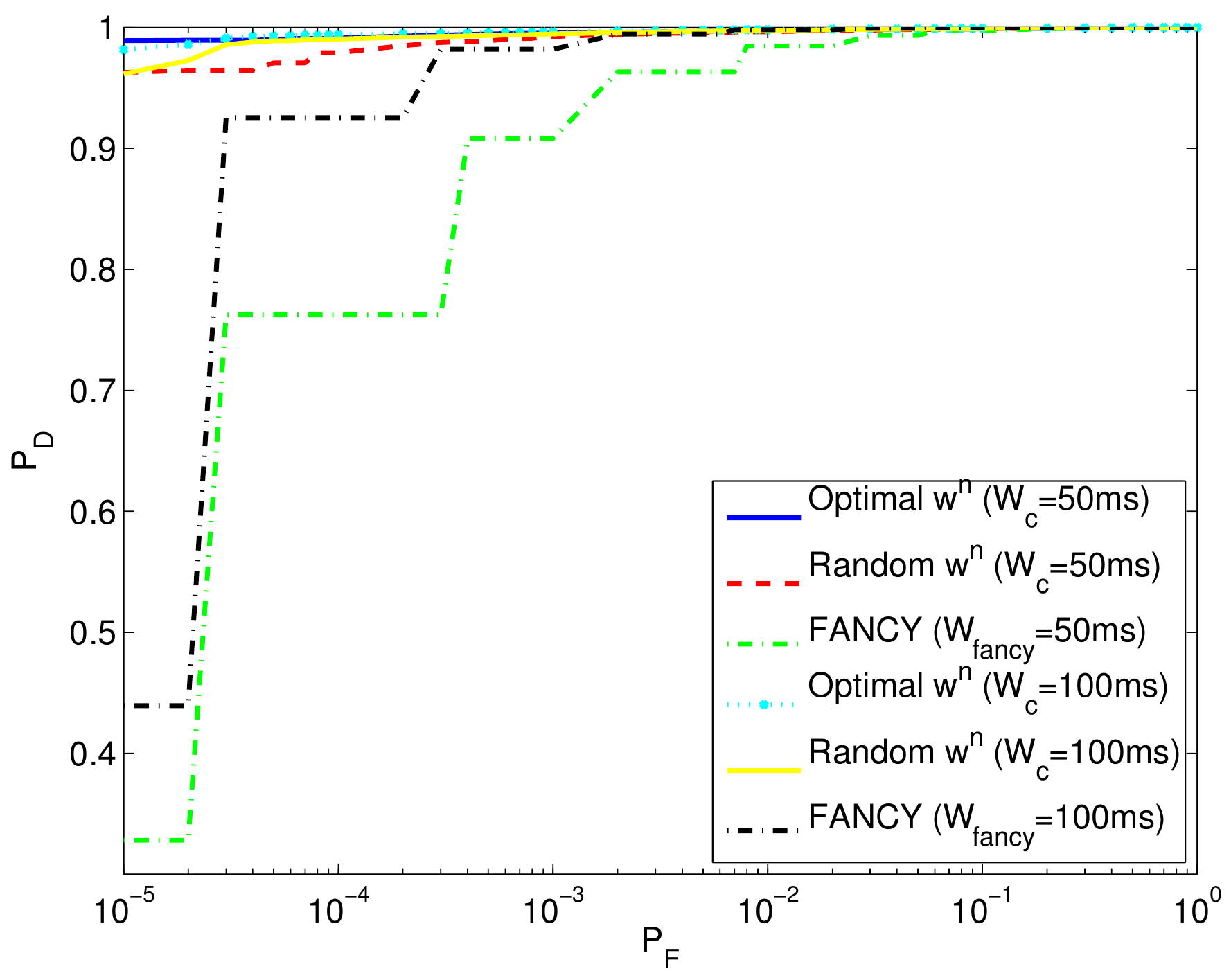}
	\caption{Scenario B ($n=20$).}
	\label{fig:Sc2FNAD} 
  \end{subfigure}
  \caption{Performance with no AD ($A_C=0$ms and $P_A=0$).}
\end{figure}

\begin{figure}
  \centering
  \begin{subfigure}[b]{\columnwidth}
	\centering
	\includegraphics[width=0.77\textwidth]{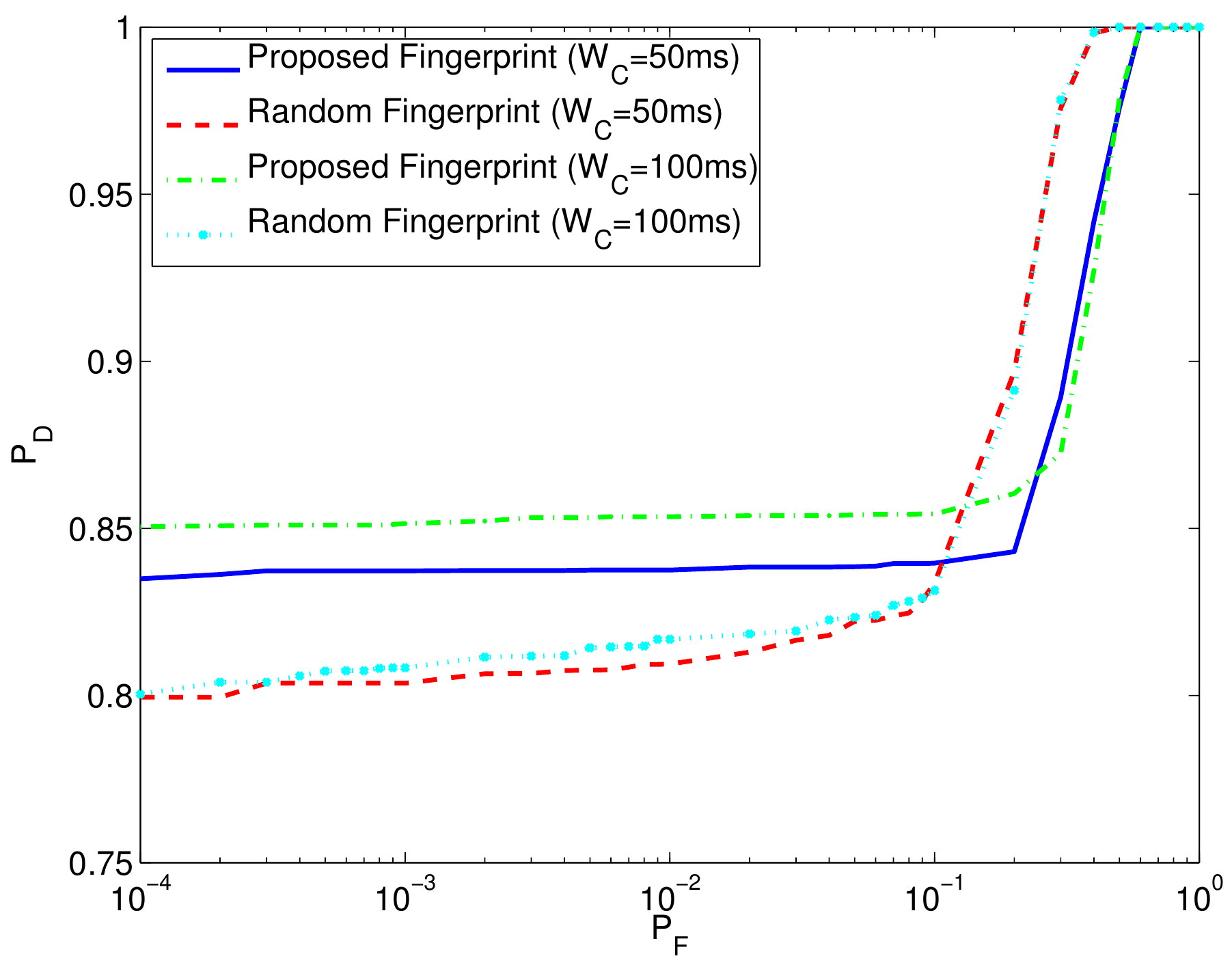}
	\caption{Scenario A ($n=20$).}
	\label{fig:Sc1FAD}
  \end{subfigure}
  \begin{subfigure}[b]{\columnwidth}
	\centering
	\includegraphics[width=0.77\textwidth]{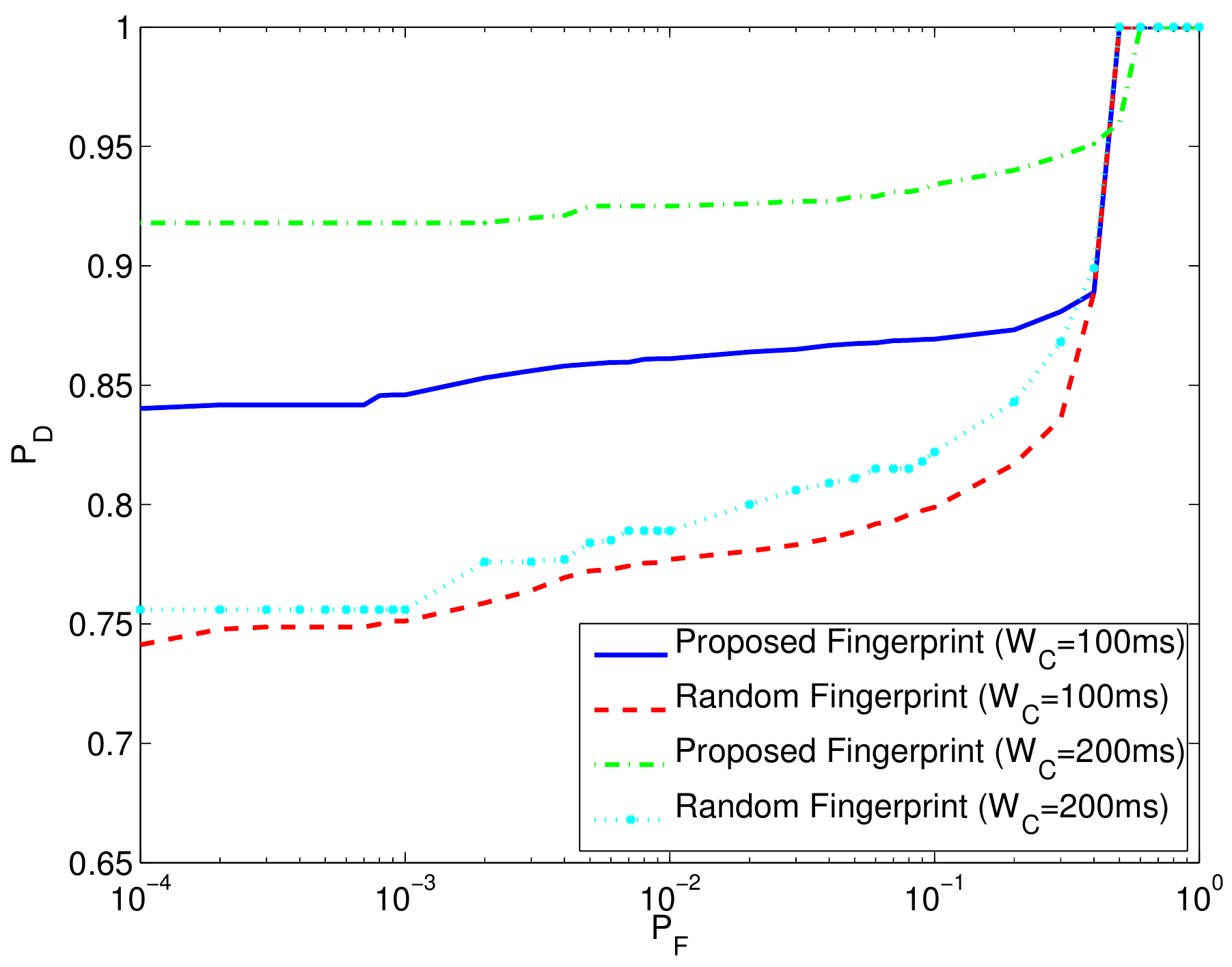}
	\caption{Scenario B ($n=30$).}
	\label{fig:Sc2FAD} 
  \end{subfigure}
  \caption{Performance with AD ($A_C=250$ms and $P_A=1$).}
\end{figure}

We compare the optimal fingerprint with two other TA-Fingerprinter strategies: a) the delays are chosen from a uniform distribution between $[0, W_C]$; b) the Fancy algorithm, that embeds its fingerprint as follows $\Delta x=\Delta u + W_{fancy} \cdot w^{n-1}$, where each $w_i \in \pm 1$, for all $i=1, \dots, n-1$, and $W_{fancy}$ is Fancy's fingerprint amplitude. We compare these strategies under the same maximum IPD variation, therefore $W_{fancy}=W_C$.
We study the performance under two conditions: 1) no AD is present ($A_C=0$ms, $P_A=0$); 2) the AD is present. In the no AD situation we use $n=5$, $W_C=1$ and 5ms in Scenario A, and $n=20$, $W_C=50$ and 100ms in Scenario B. Results are depicted in Figures \ref{fig:Sc1FNAD} and \ref{fig:Sc2FNAD} respectively. The Fancy detector outputs a sequence of bits which is error-corrected; for this reason, the ROC shows a stepwise behavior. We can see the significant difference between Fancy and the other two mechanisms, that is due to the optimality of the detector.  The difference between choosing $w^n$ optimally or randomly exists but is not so notable. In the AD environment ($A_C=250$ms and $P_A=1$), shown in  \ref{fig:Sc1FAD} and \ref{fig:Sc2FAD} for Scenarios A and B, and again the optimal fingerprint improves the performance of a uniform fingerprint, but this time the difference is more noticeable in the plots than in no AD situation. Fancy's ROC is not depicted under the AD is present conditions as it is not designed to withstand an active AD.

\section{Conclusion}\label{sec:con}
We have analyzed the flow fingerprinting game, that consists in deciding if two flows are linked or not, allowing a slight perturbation at the fingerprinter and having an adversary in the middle who tries to impair the correlation. Using this framework, we obtain the optimal detector that uses first-order statistics. Then we study the case where the adversaries' delays come from a truncated Gaussian, concluding that the adversary has to act nearly deterministically even if the detector knows the distribution. Finally, we validate the optimality of the user actions using a simulator. This simulator is also used to show that the proposed scheme outperforms the state-of-the-art in flow fingerprinting.  

\section*{Acknowledgment}
Research supported by Iberdrola Foundation through the Prince of Asturias Endowed
Chair in Information Science and Related Technologies.

\bibliographystyle{IEEEtran}

\bibliography{../bib/watermark,../bib/gametheory,../bib/traces}

\end{document}